\documentclass{aa}
\usepackage[varg]{txfonts}
\usepackage{graphicx}
\usepackage{ulem}
\usepackage{color}
\usepackage{natbib}
\bibpunct{(}{)}{;}{a}{}{,} 

\definecolor{orange}{rgb}{ 0.95, 0.60, 0}

\begin{document}

\title{A reassessment of the in situ formation of close-in super-Earths}
\titlerunning{In situ formation of close-in super-Earths}
\author{Masahiro Ogihara
\and Alessandro Morbidelli
\and Tristan Guillot
}

\institute{Observatoire de la C\^ote d'Azur, Boulevard de l'Observatoire, 06304 Nice Cedex 4, France \email{omasahiro@oca.eu}
}
\date{Received 13 February 2015 / Accepted 26 March 2015}

\abstract 
{
A large fraction of stars host one or multiple close-in super-Earth planets. There is an active debate about whether these planets formed in situ or at greater distances from the central star and migrated to their current position. It has been shown that part of their observed properties (e.g., eccentricity distribution) can be reproduced by \textit{N}-body simulations of in situ formation starting with a population of protoplanets of high masses and neglecting the effects of the disk gas.
} 
{
We plan to reassess the in situ formation of close-in super-Earths through more complete  simulations.
} 
{
We performed \textit{N}-body simulations of a population of small planetary embryos and planetesimals that include the effects of disk-planet interactions (e.g., eccentricity damping, type I migration). In addition, we also consider the accretion of a primitive atmosphere from a protoplanetary disk.
} 
{
We find that planetary embryos grow very quickly well before the gas dispersal, and thus undergo rapid inward migration, which means that one cannot neglect the effects of a gas disk when considering the in-situ formation of close-in super-Earths. Owing to their rapid inward migration, super-Earths reach a compact configuration near the disk's inner edge whose distribution of orbital parameters matches the observed close-in super-Earths population poorly. On the other hand, simulations including eccentricity damping, but no type I migration, reproduce the observed distributions better. Including the accretion of an atmosphere does not help reproduce the bulk architecture of observations. Interestingly, we find that the massive embryos can migrate inside the disk edge while capturing only a moderately massive hydrogen/helium atmosphere. By this process they avoid becoming giant planets.
} 
{
The bulk of close-in super-Earths cannot form in situ, unless type I migration is suppressed in the entire disk inside 1 AU.
}
\maketitle

\section{Introduction}
\label{sec:intro}

Recent observations of exoplanets have revealed a large number of close-in low-mass planets (e.g., \citealt{schneider_etal11}; \citealt{wright_etal11}). As of January 2015, 337 systems harbor 839 planets with masses $M < 100 M_\oplus$ (or with radii $R < 10 R_\oplus$) and semimajor axes $a < 1 {\rm AU}$ (or with orbital periods $P < 200 {\rm day}$). We define these as ``close-in super-Earths.''

They have a semimajor-axis distribution centered on 0.1 AU. Super-Earths in each system are generally confined within a few tenths of an AU. Although some of these close-in super-Earths are in or near first-order mean motion resonances (especially in 3:2 resonances), a large number of close-in super-Earths are not in mean motion resonances (e.g., \citealt{mayor_etal09}; \citealt{lissauer_etal11b}). Typical orbital separations between these planets are $10-30 r_{\rm H}$, where $r_{\rm H}$ is the mutual Hill radius, which is similar to orbital separations between the solar system terrestrial planets. Period ratios of adjacent pairs lie between the 4:3 and 3:1 resonances, respectively. It is estimated that a large number of super-Earths have small eccentricities $e\sim 0.01-0.1,$ while some of them can have high eccentricities up to $e\sim 0.5$ (e.g., \citealt{moorhead_etal11}). The mutual inclinations between planetary orbits could be estimated in a fraction of transiting systems from the Kepler catalog and appear to be low with an average of $i\sim 0.03$  \citep{fabrycky_etal14}. 

\citet{hansen_murray12,hansen_murray13} present \textit{N}-body simulations of the in situ formation of close-in super-Earths from embryos that are placed between 0.05 AU and 1 AU. To account for the masses of the known planetary systems, they had to assume that up to 100 Earth masses of solids existed in the disk within 1 AU, implying a surface density of solids much higher than for the minimum mass solar nebula (MMSN) (\citealt{weidenschilling77,hayashi81}). Assuming a nominal gas/solid density ratio of 100, this would imply a very massive protoplanetary disk that would probably be gravitationally unstable. However, \citet{chatterjee_tan14,chatterjee_tan15} propose that the disk may be enriched in solids relative to the gas thanks to the inward migration of dust grains, pebbles, and planetesimals. Thus, it may be legitimate to assume a very high surface density of solids embedded in the inner protoplanetary disk with a mass of gas comparable to gas in the MMSN model.

The simulations of \citet{hansen_murray12,hansen_murray13}, nevertheless, are quite simplistic. They start from a population of protoplanets of high masses (sometimes multiple Earth masses), which are supposed to have already achieved the completion of the oligarchic growth process \citep{kokubo_ida98}. No planetesimals are considered, and the effects of the gas, in relation to the migration and eccentricity or to the inclination damping of the protoplanets' orbits, are not taken into account. With these assumptions, their simulations suggest that an in situ accretion without gas can explain the distributions of orbital periods and eccentricities of observed super-Earths. However, their simulations did not reproduce the distribution of period ratios between adjacent super-Earths, because they had a deficit of close pairs and resonant pairs  (see Fig.~15 in \citet{hansen_murray13}).

Interestingly, the same deficit was found by \citet{cossou_etal14} in a different model accounting for the migration of planetary embryos from several AUs away. On the other hand, a model by \citet{ogihara_ida09} with gas drag and type I migration led to more separated non-resonant pairs, when type I migration was reduced by about a factor of 100.

In this paper, we revisit the process of in situ formation \citep{hansen_murray12,hansen_murray13} with more complete simulations that start from a population of small planetary embryos (i.e., Mars-mass) and planetesimals, both carrying cumulatively 50\% of the solid mass in the disk. These initial conditions are typical of terrestrial-planet simulations (e.g., \citealt{obrien_etal06}). Thus, we do not assume that the protoplanets have reached the completion of the oligarchic growth process, but we simulate that process from a much more primordial state. Moreover, we consider the action of the gas that forces a migration and eccentricity/inclination damping of the embryos and planetesimals. Our aim is to clarify differences from the studies of \citet{hansen_murray12,hansen_murray13} and to understand, with improved and more realistic simulations, whether in situ formation can explain the observed properties of the close-in super-Earth systems. We performed simulations with different disk models (i.e., the amount of solids available), but we also present a model without migration and one without gas for comparison.

In addition, we extend our model by including the accretion of primitive atmospheres onto super-Earths. The analyses of transit observations with radial velocity measurements or analyses of transit-time variations show that most of the planets larger than 2.5 Earth radii have very low densities (\citealt{marcy_etal14,hadden_lithwick14}), so are thought to have thick H/He atmospheres (up to 10-20\% by mass). We therefore consider the acquisition of H/He envelopes and discuss whether observed low-density super-Earths can be explained by our in situ accretion model.

The rest of the paper is organized as follows. In Sect.~\ref{sec:model}, we describe our model and methods of \textit{N}-body simulations; in Sect.~\ref{sec:results} we give a series of the results of our simulations; in Sect.~\ref{sec:atmosphere} we present results of simulations that include accretion of H/He atmosphere; in Sect.~\ref{sec:discussion} the conclusions are provided. 
 
\section{Model and methods}
\label{sec:model}

\subsection{Disk model}
We assume that the gas distribution is similar to the one of the classical MMSN model. Thus, for the gas surface density, we assume
\begin{eqnarray}
\label{eq:initial_disk}
\Sigma_{\rm g} = 2400 \left(\frac{r}{1 \rm{AU}} \right)^{-3/2} \exp\left(-\frac{t}{1 {\rm Myr}}\right)\,\mathrm{g\, cm}^{-2},
\end{eqnarray}
where $\Sigma_{\rm g}$ and $r$ are the gas surface density and the radial distance from the central star, respectively. The value of $\Sigma_{\rm g}$ is 1.4 times the value in the MMSN. The gas dissipation is modeled as an exponential decay with the depletion timescale of 1 Myr. We set the disk inner edge at $r = 0.1 {\rm AU}$. The inner edge of the disk is expected to be at the radius where the orbital period equals the stellar rotation period. This could be a factor of two-three smaller than we assume. Our assumption of an inner edge at 0.1 AU is dictated by constraints due to computational time. Our results concerning the position of the final planets relative to the disk's inner edge can scale with the assumed edge's position. The temperature distribution is that for an optically-thin disk \citep{hayashi81}, so that the disk scale height is
\begin{eqnarray}
\label{eq:scale_height}
h = 0.047 \left( \frac{r}{1 {\rm AU}} \right)^{5/4} {\rm AU}.
\end{eqnarray}

\subsection{Initial conditions and numerical method}
We chose an initial solid distribution similar to that of \citet{hansen_murray12}; that is, $50 M_\oplus$ in total are placed between 0.1 and 1AU. In our standard model, we set 250 embryos with a mass of $M = 0.1 M_\oplus$ and 1250 planetesimals with a mass of $M = 0.02 M_\oplus$ in such a way as to keep the radial distribution of the solid surface density proportional to $r^{-3/2}$. Planetesimals gravitationally interact with the embryos but not with each other. This set-up is typical of successful simulations for the growth of the terrestrial planets in our solar system. By adopting the same set-up, we explore the effects of the enhanced solid distribution and much shorter orbital distances.

There may be some caveats associated with the initial condition. As shown in Sect.~\ref{sec:results}, once a disk of planetesimals and planetary embryos is set, the growth of planets in the close-in region is quite rapid. If the planetesimal formation process takes a longer time than the planet-growth timescale, it would be important to take planetesimal formation into account in simulations. However, the formation of planetesimals is not yet fully understood. Therefore, in this study, we do not attempt to include the planetesimal formation; instead, we start simulations with already formed planetesimals. We note, however, that it is unlikely that planetesimal formation takes a timescale comparable to the disk lifetime ($\sim$ Myr) for two reasons. First, this timescale would correspond to 30 million orbital periods, and it is difficult to understand why it should take so long. For instance, in the solar system chondritic planetesimals formed in about one million orbits. Second, the planetesimal  formation process is most likely related to the drift of small particles through the disk \citep{johansen_etal14}, which is due to gas drag, so\ very likely most of the mass was fed to the inner disk at an early time, when the gas density was higher. A late formation of the planetesimals would require that the small particles drift into the inner part of the disk only when the disk is disappearing. We think that this possibility is difficult to envision.

\begin{table}
\caption{List of models. In Model~2, the initial solid amount is decreased by a factor of two from Model~1. Model~4 corresponds to the model of \citet{hansen_murray12,hansen_murray13}.}
\label{tbl:list}
\centering
\begin{tabular}{l l l}
\hline\hline
Model&          Type I migration&       $e,i$-damping\\
\hline 
1&                      yes&                            yes\\
2&                      yes&                            yes\\
3&                      no&                             yes\\
4&                      no&                             no\\
\hline
\end{tabular}
\end{table}

Table~\ref{tbl:list} shows the list of simulations for each model. In Model~2, the total mass inside 1AU is decreased by a factor of two (the total mass is $25 M_\oplus$). In Model~3, type I migration is neglected, but eccentricity damping is still considered. In Model~4, the effect of gas is ignored the same as in the simulation by \citet{hansen_murray12,hansen_murray13}.

Our \textit{N}-body code is based on SyMBA (Duncan et al. 1998), modified so that the effects of the gas disk are included according to the formulae reported in Sect.~\ref{sec:damping}. When bodies collide with each other, the bodies are merged by assuming perfect accretion. The physical radius of a body is determined by its mass, assuming an internal density of $\rho = 3 {\rm ~g~cm^{-3}}$. The inner boundary of the simulation is set to $r = 0.05 {\rm AU}$. We use a 0.0004-year timestep for integrations.

\subsection{Effects of gas}
\label{sec:damping}
Eccentricities, inclinations, and semimajor axes are damped by disk interaction. Planets with more than roughly $0.1 M_\oplus$ suffer the tidal damping by the density wave, while planetesimals undergo aerodynamical gas drag.

\subsubsection{Damping for embryos}
The eccentricity damping timescale for embryos, $t_e$, is given by \citep{tanaka_ward04}
\begin{eqnarray}
\label{eq:e-damp}
t_e &=& \frac{1}{0.78}\left(\frac{M}{M_*}\right)^{-1} 
\left(\frac{\Sigma_{\rm g} r^2}{M_*}\right)^{-1}
\left(\frac{c_{\rm s}}{v_{\rm K}}\right)^{4} \Omega^{-1}\nonumber\\
&\simeq& 3 \times 10^2 
\left(\frac{r}{1 {\rm AU}}\right)^2
\left(\frac{M}{M_\oplus}\right)^{-1}
\left(\frac{M_*}{M_\odot}\right)^{-1/2}
{\rm ~yr},
\end{eqnarray}
where $M_*$ is the stellar mass, $L_*$  the stellar luminosity, $c_{\rm s}$  the sound speed, $v_{\rm K}$  the Keplerian velocity, and $\Omega$  the orbital frequency, respectively. Here the relative motion between gas and planets is assumed to be subsonic ($e v_{\rm K} \lesssim c_{\rm s}$). For planets with high eccentricities and inclinations, we include a correction factor according to Eqs.~(11) and (12) of Creswell \& Nelson (2008). 

The migration timescale $t_a$ is given by \citep{tanaka_etal02,paardekooper_etal11}
\begin{eqnarray}
\label{eq:a-damp}
t_a &=& \frac{1}{\beta} \left(\frac{M}{M_*}\right)^{-1}
\left(\frac{\Sigma_{\rm g} r^2}{M_*}\right)^{-1}
\left(\frac{c_{\rm s}}{v_{\rm K}}\right)^{2} \Omega^{-1}\nonumber\\
&\simeq& 2 \times 10^5 \beta^{-1}
\left(\frac{r}{1 {\rm AU}}\right)^{3/2}
\left(\frac{M}{M_\oplus}\right)^{-1}
\left(\frac{M_*}{M_\odot}\right)^{1/2}
{\rm ~yr},
\end{eqnarray}
where $\beta$ is a coefficient that determines the direction and speed of type I migration. The type I migration torque depends on the Lindblad torque, the barotropic part of the horseshoe drag, the entropy-related part of the horseshoe drag, the barotropic part of the linear corotation torque, and the entropy-related part of the linear corotation torque. \citet{paardekooper_etal11} derived the total type I migration torque, including both saturation and the cutoff at high viscosity. We write the migration coefficient $\beta$ entering in Eq.~(\ref{eq:a-damp}) in the form
\begin{eqnarray}
\beta = \beta_{\rm L} + \beta_{\rm c,baro} + \beta_{\rm c,ent},
\end{eqnarray}
where $\beta_{\rm L},$ $\beta_{\rm c,baro},$ and $\beta_{\rm c,ent}$ are related to the Lindblad torque, the barotropic part of the corotation torque, and the entropy-related part of the corotation torque, respectively. Each formula is given by Eqs.~(11)-(13) in \citet{ogihara_etal15}. In addition, the corotation torque decreases as the planet eccentricity increases (e.g., \citealt{bitsch_kley10}). We also consider this effect using the following formula \citep{fendyke_nelson14}: 
\begin{eqnarray}
\beta_{\rm C,baro}(e) &=& \beta_{\rm C,baro} \exp\left(-\frac{e}{e_{\rm f}} \right),\\
\beta_{\rm C,ent}(e) &=& \beta_{\rm C,ent} \exp\left(-\frac{e}{e_{\rm f}} \right),
\end{eqnarray}
where $e_{\rm f} = 0.5h/r + 0.01$.

\subsubsection{Damping for planetesimals}
The aerodynamical gas drag force per unit mass is \citep{adachi_etal76}
\begin{eqnarray}
\textbf{\textit{F}}_{\rm aero} = - \frac{1}{2M} C_{\rm D} \pi r_{\rm p}^2 \rho_{\rm g} \Delta u \Delta \textbf{\textit{u}},
\end{eqnarray}
where $C_{\rm D}, r_{\rm p}, \rho_{\rm g}$, and $\Delta \textbf{\textit{u}}$ are the gas drag coefficient, the physical radius, the density of gas disk, and the velocity of the body relative to the disk of gas, respectively. Although our planetesimals have a mass of 0.02 Earth masses, we consider them as super-planetesimals, representing a swarm of much smaller objects cumulatively carrying the same total mass. From the size distribution of the asteroid belt, we think that most planetesimals had a physical size of 50km in radius \citep{morbidelli_etal09}. For the calculation of gas drag, this is the size we assume. For the value of $C_{\rm D}$, we use the same definition as in previous studies (e.g., \citealt{adachi_etal76}; Brasser et al. 2007), which depends on the Mach number, the Knudsen number, and the Reynolds number.

\section{Results}
\label{sec:results}

\subsection{Outcomes of Models~1 and 2}
\begin{figure}
\resizebox{1.0 \hsize}{!}{\includegraphics{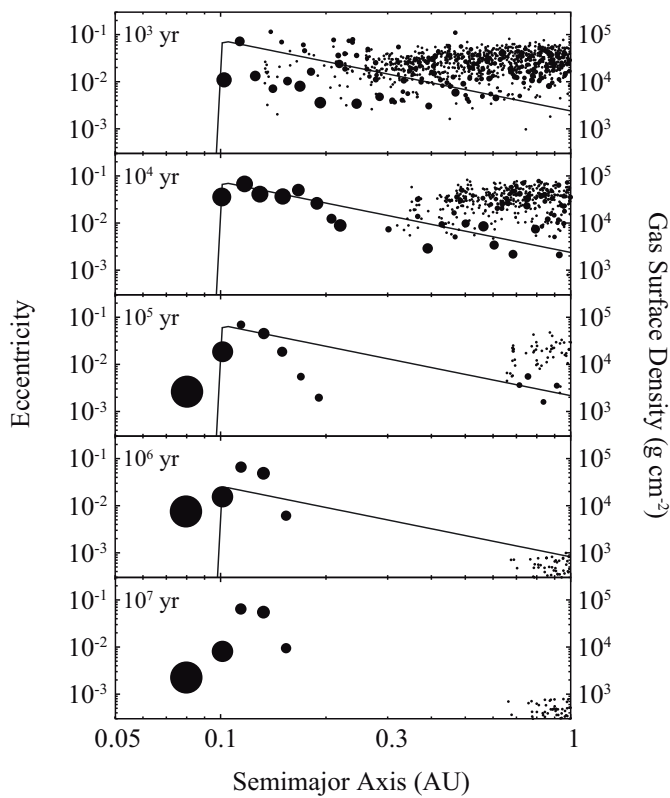}}
\caption{Snapshots of a system for Model~1. Filled circles represent bodies. The size of the circles is proportional to the radius of the body. The smallest circle represents a 0.02 Earth-mass body, while the largest one represents a 33 Earth-mass body. The solid line indicates the gas surface density (right axis).
}
\label{fig:snap}
\end{figure}

\begin{figure}
\resizebox{1.0 \hsize}{!}{\includegraphics{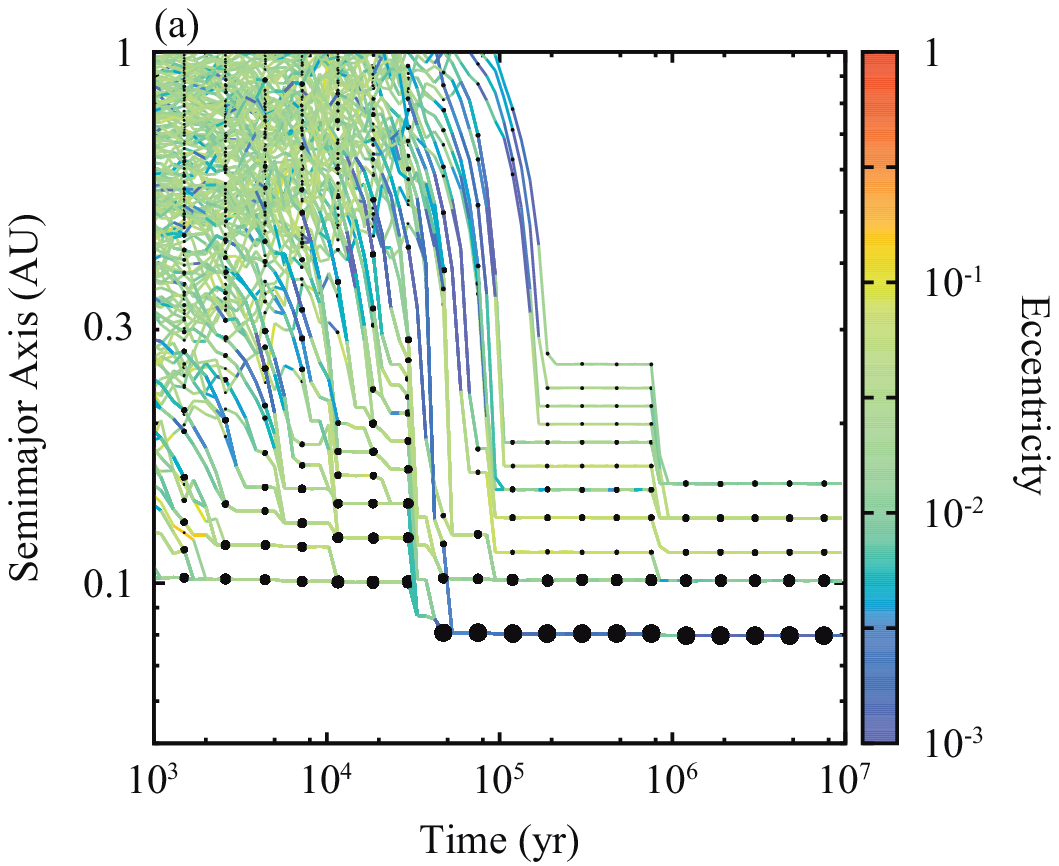}}
\resizebox{1.0 \hsize}{!}{\includegraphics{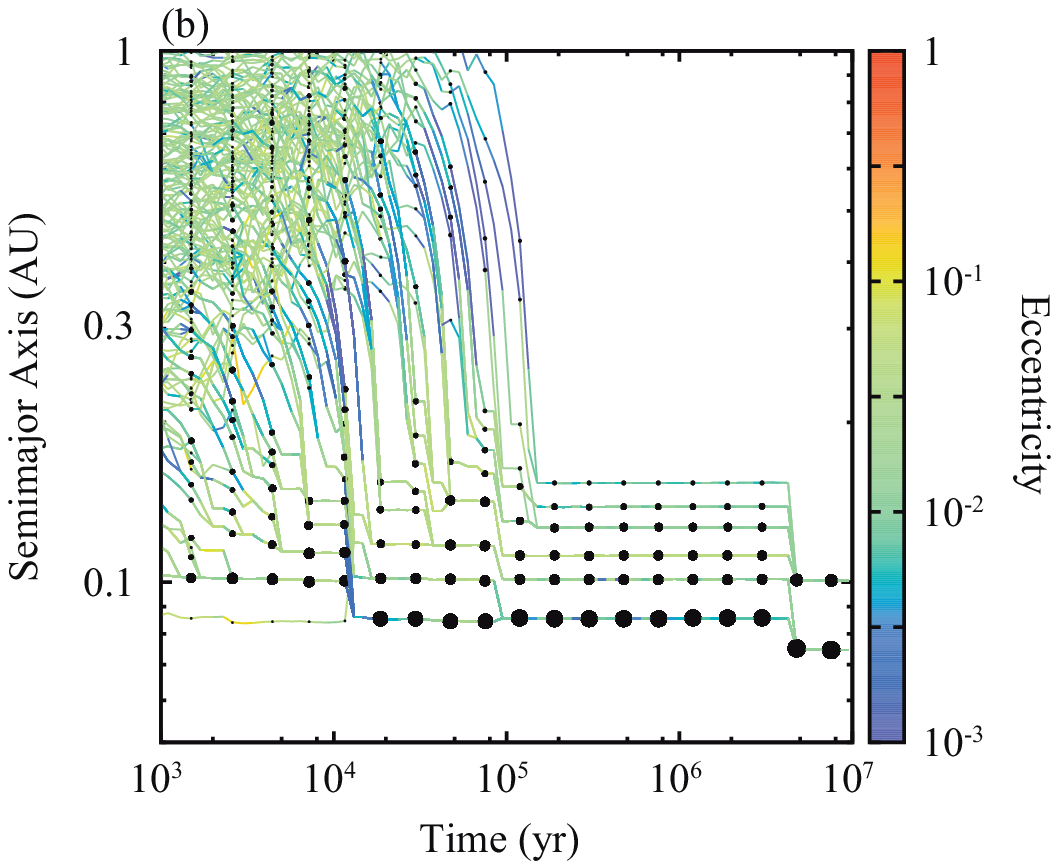}}
\caption{Time evolution of planets for Model~1. The filled circles connected with solid lines represent the sizes of planets. The smallest circle represents a 0.2 Earth-mass embryo, while the largest ones represent a 33 Earth-mass planet in panel~(a) and 35 Earth-mass planet in panel~(b). The color of line indicates the eccentricity (color bar).
}
\label{fig:t-a}
\end{figure}

The formation of close-in super-Earth systems can be schematically divided into three phases: (i) growth of embryos from planetesmals, (ii) migration of embryos, and (iii) gas depletion and long-term orbital evolution. Through a series of simulations, we find that the first accretion stage is extremely rapid ($\sim$0.1 Myr, much faster than the typical timescale of several 10 Myr observed in terrestrial planet simulations). Thus, the effect of the enhanced amount of solid is not just that of producing bigger planets: it also reduces the formation timescale. The closer proximity to the central star (i.e., shorter orbital periods) also favors a much faster accretion (see also \citealt{lee_etal14}). Figure~\ref{fig:snap} shows snapshots of the evolution of one simulation for Model~1and indicates the gas surface density at each time (right axis). Figure~\ref{fig:t-a}(a) shows the time evolution of the semimajor axis for this run. The color of lines indicates the eccentricity of the planets (see color bar). At $t = 10^3 {\rm yr}$, almost all planetesimals initially placed inside $r \simeq 0.2 {\rm AU}$ have been accreted by embryos. The first accretion phase ends before $t=0.01 {\rm Myr}$ and $0.1 {\rm Myr}$ inside $r\simeq 0.4 {\rm AU}$ and $\simeq 0.7 {\rm AU}$, respectively.

Because of the short accretion timescale, it is thus not correct to neglect the gas effects as in Hansen and Murray's works. In fact, the protoplanets become massive well before the gas disk is substantially depleted. The gas forces the planets to migrate inward. All planets would be lost into the star if there were no inner edge of the disk. With a sharp disk inner edge, the innermost planet is trapped at the edge by the planet-trap effect \citep{masset_etal06}. The other planets pile up in mutual mean motion resonances with the former. The typical commensurabilities are between 5:4 and 6:5 with orbital separations of $\simeq 5-10 r_{\rm H}$, thus the final orbits are packed near the disk's edge. The resonant configurations of two bodies that undergo convergent migration are determined by the mass of the bodies and the relative migration speed (e.g., \citealt{mustill_wyatt11}; \citealt{ogihara_kobayashi13}). \citet{ogihara_kobayashi13} have derived the critical migration timescale for capture into first-order mean motion resonances. They found that if the relative migration timescale is shorter than $t_{a,{\rm crit}} \simeq 1 \times 10^5 (M_1/M_\oplus)^{-4/3} T_{\rm K}$, where $M_1$ is the mass of larger body and $T_{\rm K}$  the Keplerian period, bodies can only be captured in resonances closer than the 4:3 resonance (see Eq.~(6) and Table~2 in \citealt{ogihara_kobayashi13}). In the result of Model~1, the mass of migrating embryos is $\simeq 0.5 M_\oplus$ and the mass of planets near the edge is $\sim 1 M_\oplus$, so the relative migration timescale is $\simeq 10^5 T_{\rm K}$ (Eq.~(\ref{eq:a-damp})). Migrating embryos therefore settle into closely packed configurations. Five planets form at the end of the simulation presented in Figs.~\ref{fig:snap} and \ref{fig:t-a}(a). In this run, the planets do not exhibit orbital instability after gas depletion because the number of planets in the system is small and the planets are in mean motion resonances, leading to a long-lasting orbital stability (\citealt{chambers_etal96}; \citealt{matsumoto_etal12}).

\begin{figure}
\resizebox{0.9 \hsize}{!}{\includegraphics{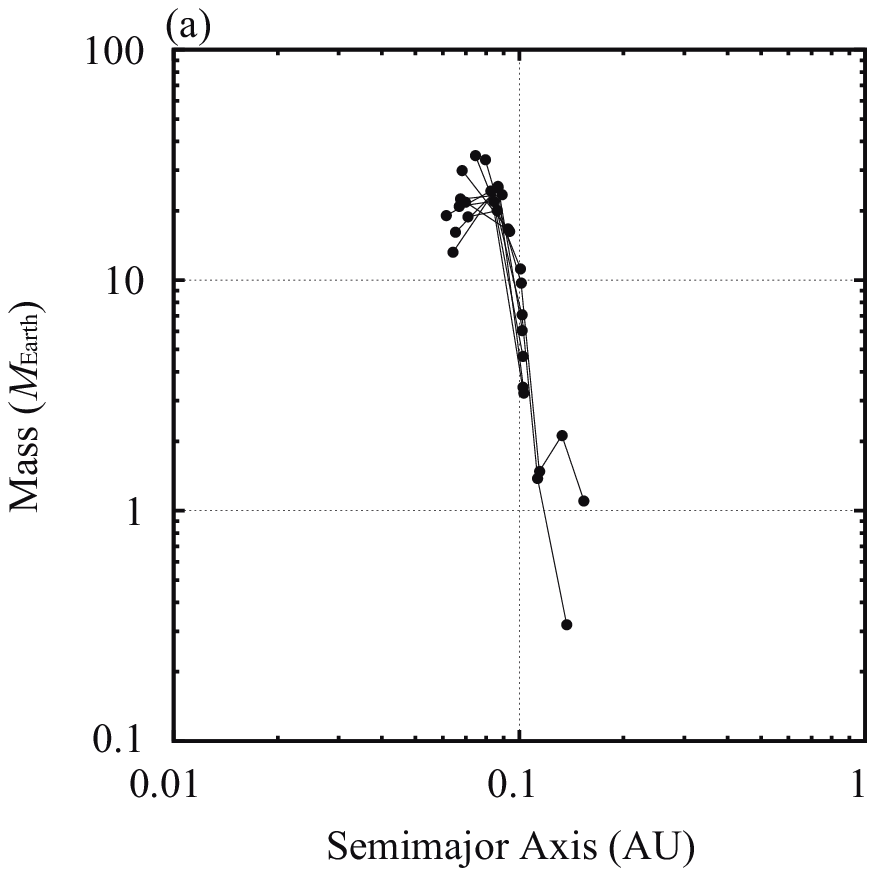}}
\resizebox{0.9 \hsize}{!}{\includegraphics{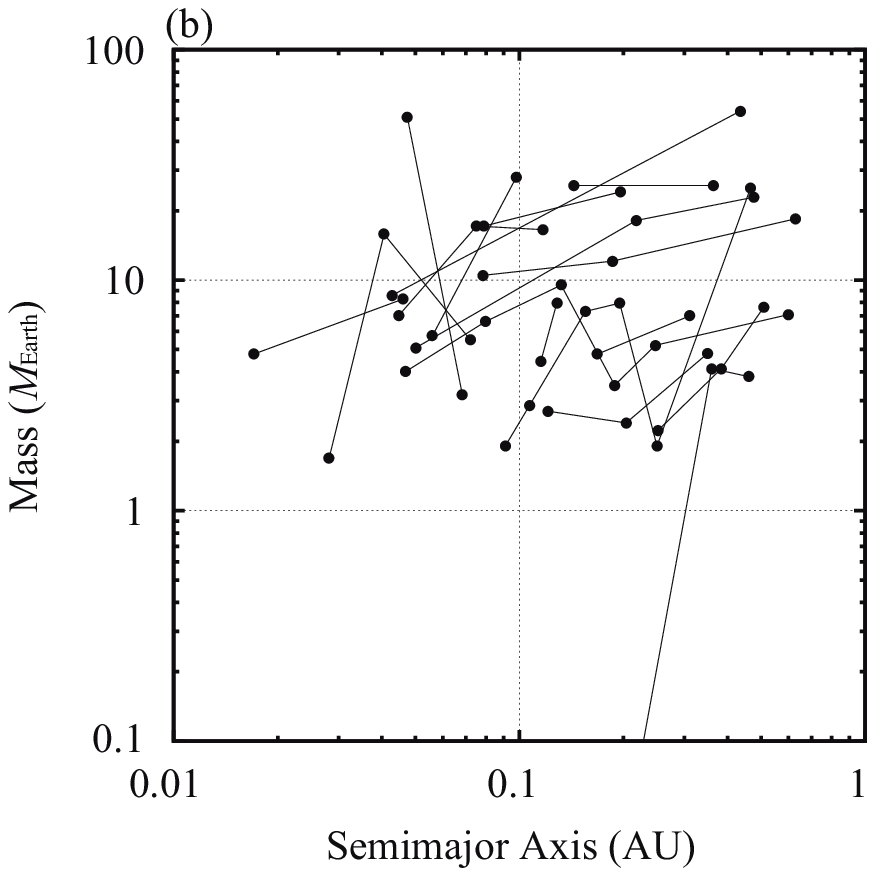}}
\caption{Results of 10 simulations of Model~1 (panel (a)). Observed close-in super-Earths systems (panel (b)).
}
\label{fig:a-m}
\end{figure}

We performed ten simulation runs for each model with different initial positions of embryos and planetesimals. The results are qualitatively the same: final orbits are compact near the disk's inner edge. In some runs, the system undergoes orbital instability during the third phase after 1~Myr, resulting in non-resonant and relatively separated orbits with a smaller number of planets (see Fig.~\ref{fig:t-a}(b) for example). Figure~\ref{fig:a-m}(a) shows the final orbital configurations of all ten runs, where the planets that formed through the same run are connected with a line. We observe steep mass gradients in this figure; the largest bodies are located near the edge, and the planetary mass monotonically decreases when the semimajor axis increases. This is because, in the presence of a strong migration torque, the resonant system is stable only if the innermost planet is the most massive, as already found by \citet{morbidelli_etal08}. If originally the innermost planet is not the most massive, an instability typically occurs. The first and the second planets have encounters with each other, and the system stabilizes in resonance only after that the two planets have exchanged their relative positions. The same is true for the second, relative to the third planet and so forth.

Figure~\ref{fig:a-m}(b) shows orbital configurations of observed close-in super-Earth systems, in which the mass and semimajor axis of all planets are known. It is clear that the bulk architecture of the observed systems is inconsistent with the simulated planetary systems. No steep mass gradients are observed\footnote{The mass gradient is steep in the Kepler-101 system, where the innermost planet has 51 Earth mass and the second planet is about 3 Earth mass.}, and the bulk architecture of the observed systems cannot be reproduced through the simulations of model~1. If embryos migrate from beyond 1 AU, the mass gradient would be shallower and/or the outer boundary of the final planet distribution at around 0.2 AU would be removed. This would suggest that a ``migration model'' yields better results in reproducing the observed close-in super-Earth systems, which should be investigated by future \textit{N}-body simulations.

\begin{figure*}
\begin{center}
\resizebox{0.7 \hsize}{!}{\includegraphics{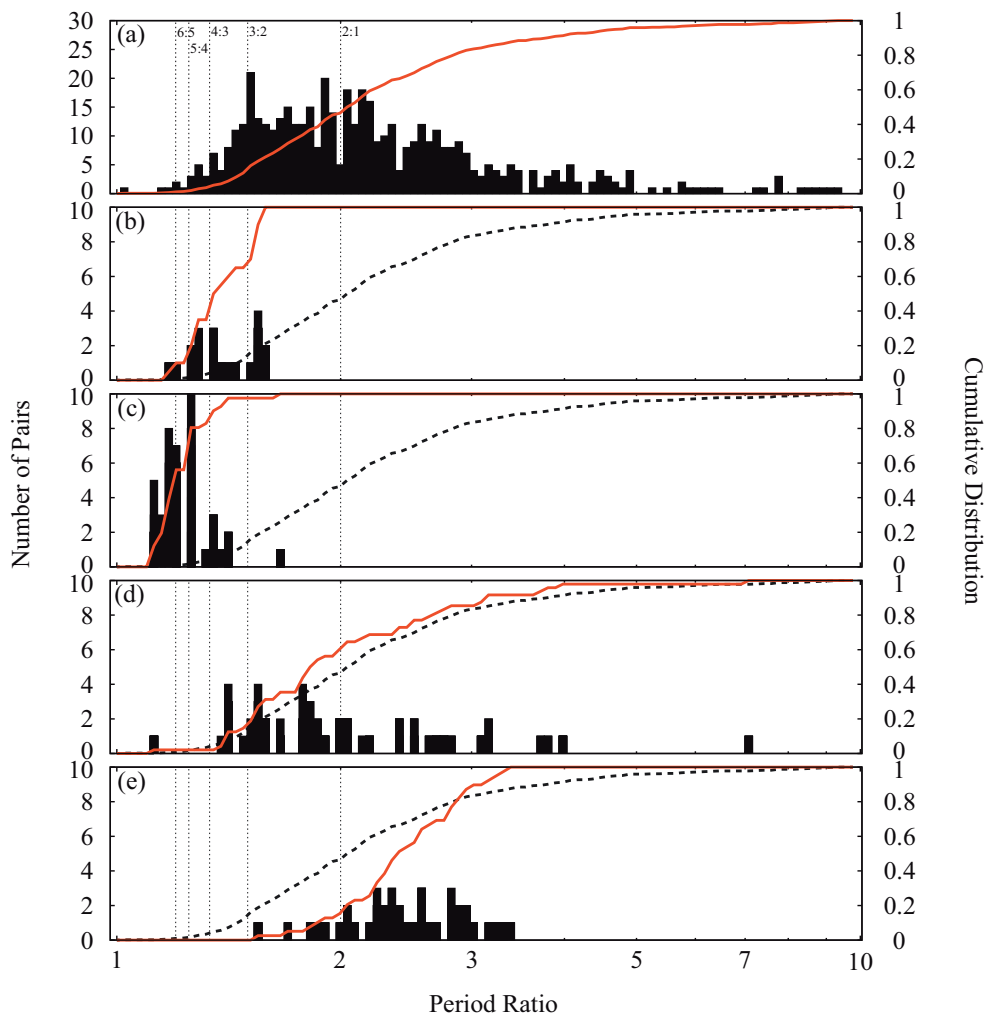}}
\end{center}
\caption{Comparison of the distributions of period ratios of adjacent pairs of planets for (a) observation and (b)-(d) simulations. The distribution of period ratios is presented as a histogram (see the left y-axis) and as a cumulative distribution (see the right y-axis). Panels~(b), (c), (d), and (e) show the results of Models~1, 2, 3, and 4, respectively. The dashed lines in panels~(b)-(d) represent the cumulative distribution of observed planets shown in red in panel (a). The vertical lines indicate locations of mean motion resonances.
}
\label{fig:p-n}
\end{figure*}

Figure~\ref{fig:p-n}(a) shows period ratios of adjacent pairs of observed close-in super-Earths indicating period ratios of first-order mean motion resonances (e.g., 2:1 and 3:2) and the cumulative distribution of the period ratios (right axis). Most pairs have period ratios between 4:3 and 3:1, and there are a few pairs that have period ratios lower than 4:3. Figure~\ref{fig:p-n}(b) shows the period ratio distribution for the ten runs for Model~1 and a copy of the observed distribution. Some pairs are in closely-spaced resonances (e.g., 5:4), and others have been knocked out of resonant orbits during the late instability. Although very closely spaced pairs ($\lesssim$ 4:3) can form in Model~1, the cumulative distribution of observed close-in super-Earths is not matched by the results of Model~1. In fact, a Kolmogorov-Smirnov (K-S) test indicates that the cumulative distributions are statistically different $(Q_{\rm KS} \ll 0.01)$. 

\begin{figure}
\resizebox{1.0 \hsize}{!}{\includegraphics{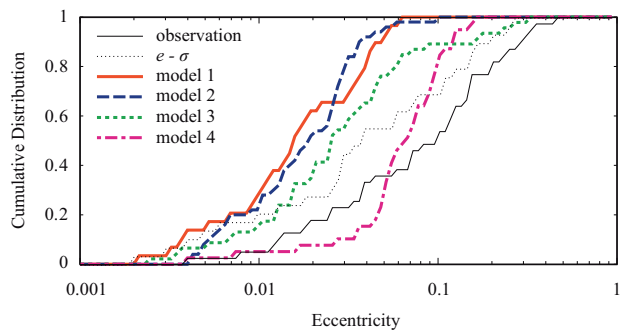}}
\caption{Comparison of cumulative eccentricity distributions between observed close-in super-Earths (thin solid line) and the planets formed through simulations (thick lines, see legend). The thin dotted line indicates the cumulative eccentricity distribution of observed super-Earths, in which each eccentricity is assumed to be $e - \sigma$.
}
\label{fig:e-n}
\end{figure}

Figure~\ref{fig:e-n} shows cumulative distributions of the eccentricities of the observed close-in super-Earths and of all the results produced in the simulations corresponding to the same model. The general trend is that the eccentricity of the results of Model~1 is smaller than for close-in super-Earths. In  Model~1, planets with $e < 0.03$ account for 66 percent of all bodies, while exoplanets with $e < 0.03$ make up only 26 percent of all planets. One reason for the small eccentricity in Model~1 is that the number of planets in a system is small $(N = 3-6),$ and the orbital stability time is long \citep{chambers_etal96}, which inhibits giant impacts during gas dispersal (see Fig.~\ref{fig:t-a}(a) for example). In addition, even if planets undergo orbital instability as in Fig.~\ref{fig:t-a}(b), the eccentricity is not highly excited. This is because the effect of mutual scattering between planets is limited by the small number of planets. Eccentricity damping also operates during the gas dissipation phase due to remnant gas (see Fig.~\ref{fig:t-a}(b), for example).
A K-S test indicates that the observed and simulated eccentricity distributions are statistically different $(Q_{\rm KS} \ll 0.01)$. 
In summary, the results of Model~1 cannot reproduce the bulk properties (period ratio, eccentricity) of observed close-in super-Earth systems. Eccentricities of exoplanets can be overestimated (see Sect.~\ref{sec:results_model3} for discussion).
Thus, our results disagree with those of Hansen and Murray, which is not surprising given that the latter neglected the effect exerted by the disk of gas (particularly the inward migration of proroplanets), not realizing that the super-Earths must form well within the gas-disk lifetime.

We also performed ten runs for Model~2, where the initial solid amount is reduced by a factor of two from Model~1, in the hope of observing a slower accretion rate and consequently weaker migration effects. However, the results are qualitatively the same with those of Model~1. In Fig.~\ref{fig:p-n}(c), planetary pairs with their period ratio of $<$ 4:3 account for 78 percent of all pairs, and the cumulative distribution is different from that of observed close-in super-Earths. The results are even more closely packed than those of Model~1 because planets are less vulnerable to orbital instability during the third phase. The eccentricity distribution in Fig.~\ref{fig:e-n} also differs from that of close-in super-Earths.

\subsection{Outcomes of Models~3 and~4}
\label{sec:results_model3}
We then present results of Models~3 and~4, in which we suppress the migration torques exerted by the gas-disk interaction onto the planets (Model~3) or we neglect the presence of gas altogether (Model~4). Clearly, both models are academic. In fact, a general mechanism that suppresses type I migration has never been found. Locally, type I migration can be halted or reversed \citep{paardekooper_mellemal06,bitsch_etal14}, but no global weakening of type I migration has even been demonstrated. As for the absence of gas, this seems inconsistent with the fast growth timescale for the super-Earths. It is difficult to imagine that the gas disappears significantly faster than what we assume above (1 Myr). The reason we present these models is to highlight the role of migration and eccentricity/inclination damping in the results we presented before.

\begin{figure}
\resizebox{1.0 \hsize}{!}{\includegraphics{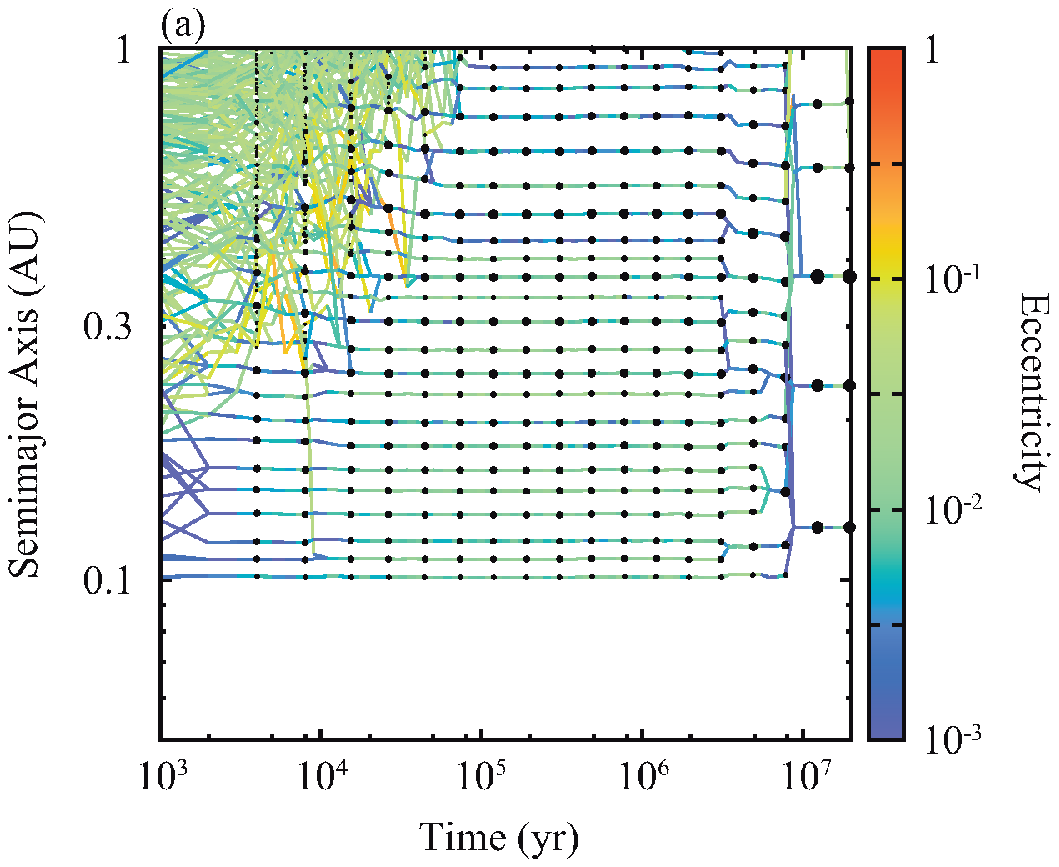}}
\resizebox{1.0 \hsize}{!}{\includegraphics{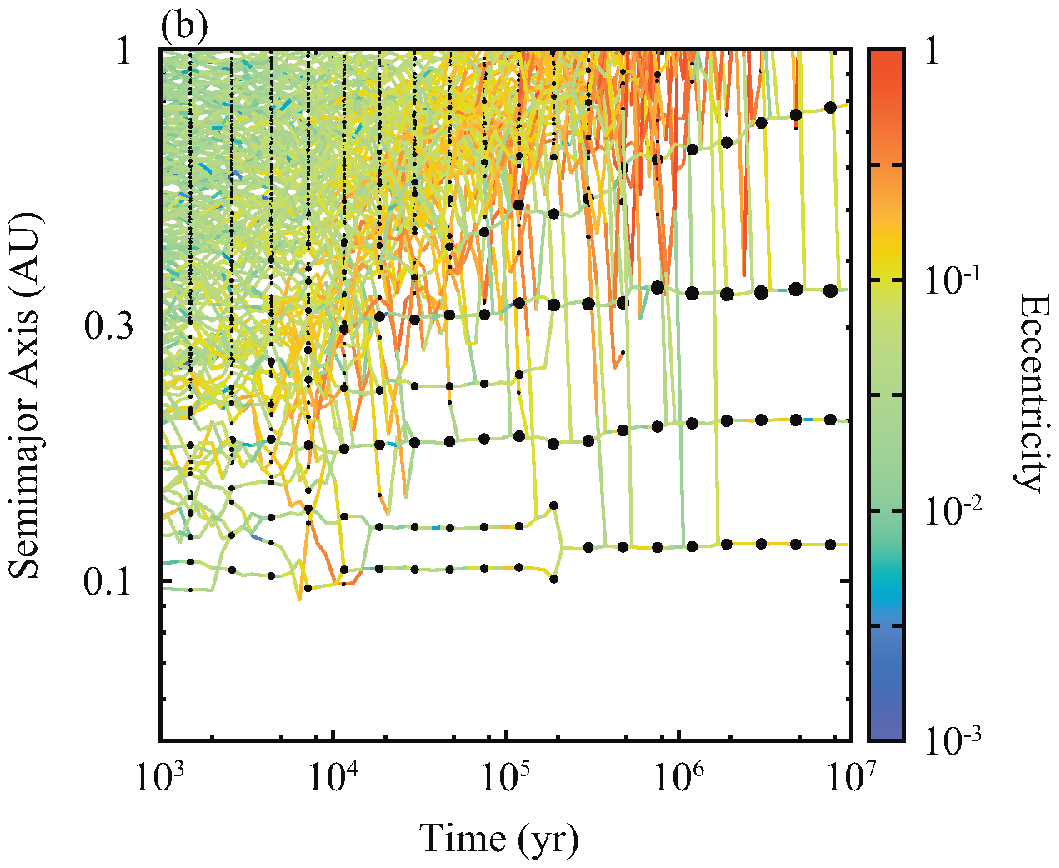}}
\caption{Same as Fig.~\ref{fig:t-a} but for a representative simulation of Model 3 (panel (a)) and Model (4) (panel (b)).
}
\label{fig:t-a-others}
\end{figure}

Figures~\ref{fig:t-a-others}(a) and (b) show the typical orbital evolution for Models~3 and~4, respectively. In both cases, planetary systems are not as compact when compared with the results of Models~1 and 2. In the simulations for Model~3, in which type I migration is neglected, planets undergo slow inward migration due to eccentricity damping (see also Sect.~5.2 in \citealt{ogihara_etal14}), and planets are temporarily captured in mutual mean motion resonances before a few Myr. Then, they undergo orbital instability and collide with each other, resulting in non-resonant orbital configurations, which are qualitatively similar to those obtained in the slow-migration simulation of \citet{ogihara_ida09}. In the results for Model~4, planets are never in resonances, which is almost the same as in the simulations of \citet{cossou_etal14} and \citet{hansen_murray12,hansen_murray13}. The averaged mass of the largest bodies is $\simeq 13~ M_\oplus$ (Model~3) and $\simeq 14~ M_\oplus$ (Model~4), which are lower than for Model~1.

Figures~\ref{fig:p-n}(d) and (e) show the period ratio distribution of Models~3 and 4, respectively. Interestingly, the results match the observations much better than those of Models~1 and~2. In the results of Model~3, most pairs have period ratios between 4:3 and 3:1, while some pairs are relatively separated (> 3:1). In the results of Model~4, 85 percent of pairs lies between 2:1 and 3:1. In comparison with the observed distribution, Model~3 is a good match to the distribution of exoplanets. A K-S test shows that the distributions are similar with a significance level of $Q_{\rm KS} = 0.24$ for Model~3, while the distribution of Model~4 is not similar to the observed distribution $(Q_{\rm KS} \ll 0.01)$. 

In Fig.~\ref{fig:e-n}, the eccentricity of the planets produced in Model~3 is generally smaller than in Model~4 because of the eccentricity damping. Compared with the distribution of exoplanets, the K-S test shows that the distributions are different for Model~3 $(Q_{\rm KS} \ll 0.01),$ 
while the distributions for Model 4 are closer, but are still not satisfactory $(Q_{\rm KS} = 0.026)$. Thus, neither Model~3 nor Model~4 explains the observation, because the first has problems with the eccentricity distribution, the second with the orbital period distribution.

However, it is fair to say that the measurement of the eccentricity of exoplanets is still difficult, and the uncertainty on the results is quite large. In particular, it has been shown that eccentricities, which are derived from radial velocity surveys, can be overestimated (e.g., \citealt{shen_turner08}; \citealt{zakamska_etal11}). Therefore, the observed eccentricity distribution in Fig.~\ref{fig:e-n} may shift to lower values. In this case the results of Model~3 might match the eccentricity distribution as well.

As an example, we recalculated the eccentricity distribution in a simple way. The eccentricity of each exoplanet is set to $e - \sigma$, where $\sigma$ is the estimated error. The new distribution is indicated in Fig.~\ref{fig:e-n}. We find that the new distribution gives a better match to Model~3 $(Q_{\rm KS} = 0.30)$ rather than Model~4 $(Q_{\rm KS} = 0.010)$. 

\section{Accretion of primitive atmospheres}
\label{sec:atmosphere}

An objection often expressed against the in situ accretion model for super-Earths is that observations indicate in many cases (e.g., Kepler-11, see \citealt{lissauer_etal11a}) that these planets have a low bulk density. One would expect planets grown in the inner part of the disk to be rocky, given that the high local disk temperature should not have allowed ice condensation.

A possibility, however, is that super-Earths accreted primitive H/He atmospheres, leading to low bulk densities. Recent simulations of the structure and evolution of planetary atmospheres have demonstrated that super-Earths can indeed accrete primordial atmospheres from the protoplanetary disk provided that they do not accrete solids at a high rate \citep{lee_etal14}. As we have seen above, we expect that in situ formation is extremely rapid and reaches completion well within the lifetime of the gas disk. Thus, we expect our planets to be in the condition of a very low accretion rate of solids when there is still gas in the disk, enabling the acquisition of significant atmospheres. In what follows we implement the most recent recipes for gas accretion to compute the mass of atmosphere expected for our super-Earths.

\subsection{Model}
\label{sec:atmos_model}
Once planetary embryos embedded in a gas disk are sufficiently massive (typically the mass of Mars or more), they capture part of the disk gas to have an atmosphere of their own \cite[e.g.,][]{wuchterl+2000}. This atmosphere or envelope grows with the mass of the embryo itself until it becomes greater than the mass of the solid core. At this so-called crossover mass ($M_{\rm env}\sim M_{\rm core}$), accretion enters a runaway phase with the envelope mass increasing exponentially and the planet becoming a giant planet \citep{pollack+1996}. In this process, the accretion of solids has two effects: it increases the mass of the planet and heats the envelope. The first favors the growth of the envelope, but the second leads to an increase in the crossover mass because of a more tenuous envelope. Most works on the growth of giant planet cores have thus focused on obtaining expressions that depend on the accretion rate of planetesimals \citep[e.g.,][]{ikoma_etal01}.

However, since in our simulations the accretion of solids stalls while the gas disk is still present, at this later stage, the rate of cooling of the envelope becomes crucial for controlling the growth of the envelope \citep{pollack+1996}. We choose to empirically model the growth of the envelope by fitting the results of \cite{ikoma_hori12}, \cite{piso_youdin14}, and \cite{lee_etal14}, which are works that account for the planetary cooling to calculate the resulting envelope growth. Specifically, we model the envelope mass, $M_{\rm env}$, as
\begin{equation}
\frac{M_{\rm env}}{M_{\rm core}}=\frac{k_2}{1+k_3} \left[ 1+k_3 \left(\frac{t}{t_{\rm run}}\right)^{1/3}\right]e^{k_1 (t/t_{\rm run}-1)},
\label{eq:Menv}
\end{equation} 
where $M_{\rm core}$ is the core mass and $t_{\rm run}$  the time to get to the crossover mass. As for the coefficients, we use $k_1 = (M_{\rm core} /15 M_\oplus)^{-1},$  $k_2 = 1 + (M_{\rm core} / 40 M_\oplus)$, and $k_3= 9$. Most of the uncertainty is linked to the value of $t_{\rm run}$, which depends critically on the opacities chosen, the cooling rate of the core, and orbital distance. For the present simulations we choose to use as a fiducial value
\begin{equation}
t_{\rm run}=10^{7} (M_{\rm core}/5 M_\oplus)^{-3}\rm\ yr,
\end{equation}
which lies in between the results obtained by \citet{ikoma_hori12} and  \citet{piso_youdin14} and  \citet{lee_etal14}.

Then the accretion rate onto the planet is expressed by
\begin{equation}
\dot{M}_{\rm env} = M_{\rm env}\left[\frac{k_3}{3}\frac{1}{t_{\rm run}^{1/3}t^{2/3}+k_3 t}+\frac{k_1}{t_{\rm run}}\right].
\label{eq:mdot_env}
\end{equation}
Equation~(\ref{eq:Menv}) implicitly assumes that the disk can supply all the gas that the planet is able to accrete. In reality, this amount is limited by the viscous inflow in the disk at the location of the planet \citep[e.g.,][]{tanigawa_ikoma07}. The accretion rate due to viscous diffusion is
\begin{equation}
\dot{M}_{\rm vis} \simeq 3 \pi \nu \Sigma_{\rm g},
\label{eq:mdot_vis}
\end{equation}
where an ``alpha model'' for the disk viscosity is used with $\nu = \alpha c_{\rm s} h$ ($c_{\rm s}$ is the isothermal sound speed), and we adopt $\alpha = 10^{-3}$. The actual accretion rate is limited by the minimum of Eqs.~(\ref{eq:mdot_env}) and (\ref{eq:mdot_vis}).

Based on these prescriptions, we recalculate our \textit{N}-body simulations for Model~1, this time accounting for the accretion of an atmosphere after the end of core accretion phase ($t \lesssim 10^5 {\rm yr}$).

\subsection{Results}
We now present the results of these simulations with the set-up of Model~1 but accounting for atmosphere accretion. There are two crucial questions that we wish to address: 1) Can atmosphere accretion be substantial enough to significantly reduce the apparent bulk density of the planet and mimic low-density objects as in the Kepler-11 system? 2) Can atmosphere accretion change the masses of the planets enough to induce a late dynamical instability that can result in less compact systems.

\begin{figure}
\resizebox{1.0 \hsize}{!}{\includegraphics{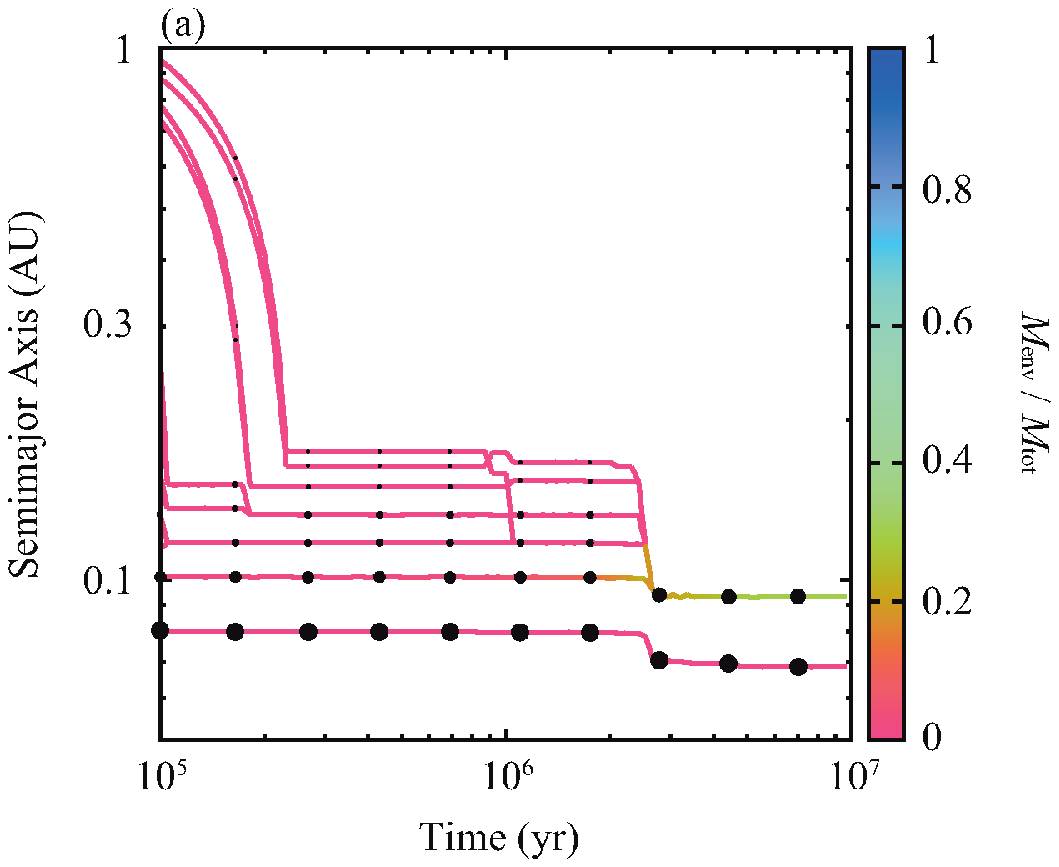}}
\resizebox{1.0 \hsize}{!}{\includegraphics{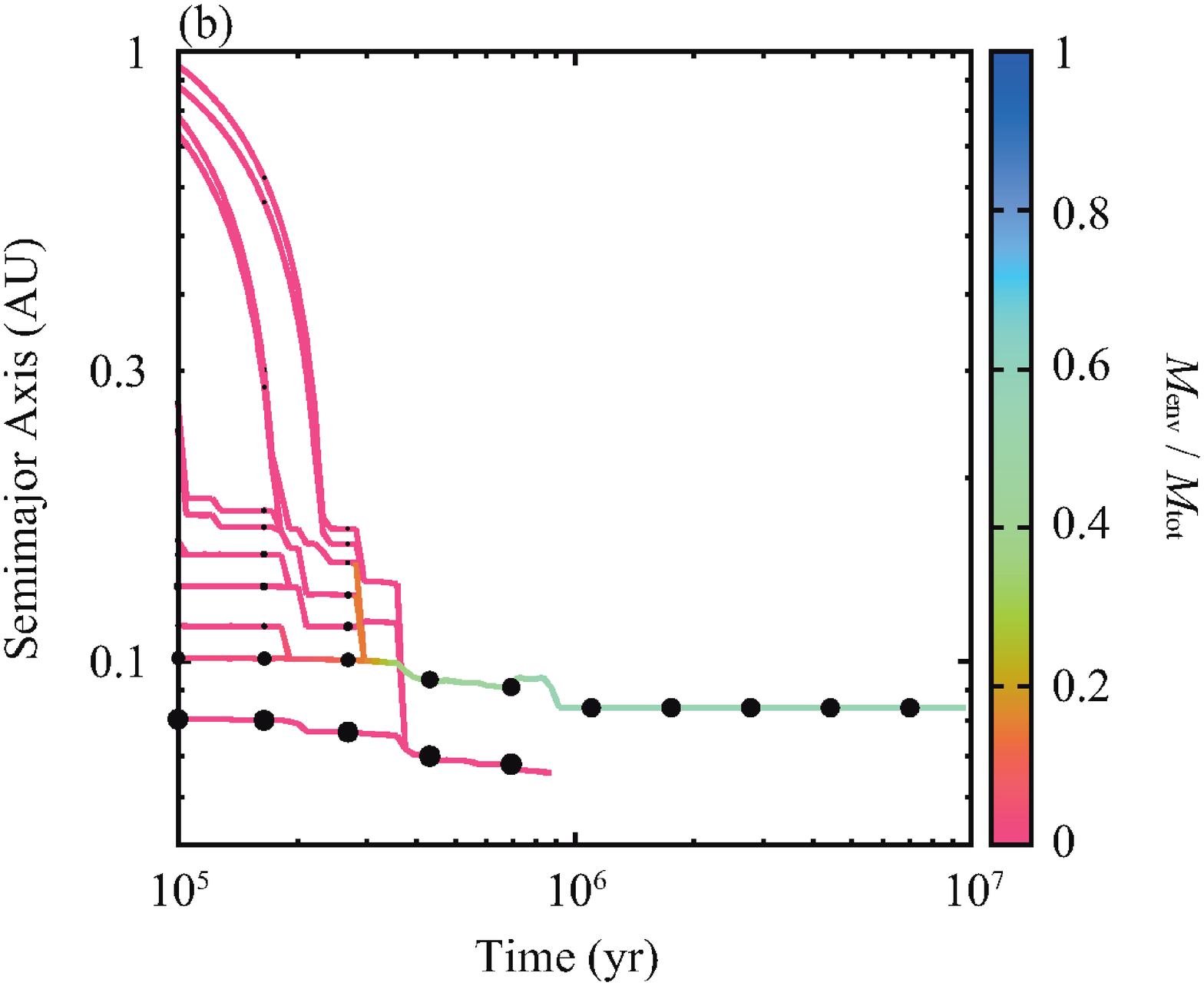}}
\caption{Evolution of the semimajor axis and the envelope mass (color bar). Panel (a) shows the result of $t_{\rm run} = 10^{7} (M_{\rm core}/5 M_\oplus)^{-3}\rm\ yr$, while panel (b) indicates that of $t_{\rm run} = 10^{6} (M_{\rm core}/5 M_\oplus)^{-3}\rm\ yr$.  The filled circles connected with solid lines represent the sizes of planets. The largest circles represent a 33 Earth-mass planet in panel~(a) and 36 Earth-mass planet in panel~(b).
}
\label{fig:t-a-env}
\end{figure}

Figure~\ref{fig:t-a-env} shows the orbital evolution of each system in which the color of each planet's lines indicates the ratio of envelope mass to total mass, $M_{\rm env}/M_{\rm tot}$ (see color bar). We adopt  $t_{\rm run}=10^{7} (M_{\rm core}/5 M_\oplus)^{-3}\rm\ yr$ as a fiducial value for $t_{\rm run}$ in the result of panel (a). In addition, we also perform a simulation under more efficient conditions for envelope accretion $t_{\rm run}=10^{6} (M_{\rm core}/5 M_\oplus)^{-3}\rm\ yr$, the results of which are shown in panel (b).

In Fig.~\ref{fig:t-a-env}(a), the planet at $a=0.1 {\rm AU}$ with $M_{\rm core} = 9.2 M_\oplus$ accretes gas from the disk and ends up retaining a thick atmosphere of $M_{\rm env} = 6.2 M_\oplus$ and $M_{\rm env}/M_{\rm tot} = 0.30$ at $t = 10 {\rm Myr}$. This planet migrates inside of the disk inner edge at $t \simeq 2.6 {\rm Myr}$ by the interaction with four outer bodies. This migration prevents the former planet from accreting more gas. The innermost planet is also moved inside the disk inner edge before $t=0.1 {\rm Myr}$ and stops accreting gas at that point.

In the simulation with a shorter value of $t_{\rm run}$ shown in Fig.~\ref{fig:t-a-env}(b), a planet with $M_{\rm env} = 14 M_\oplus$ and $M_{\rm env}/M_{\rm tot} = 0.56$ eventually forms and also moves inside of the disk edge at $t \simeq 0.3 {\rm Myr}$. It is interesting to notice that as the envelope mass increases, systems become vulnerable to orbital instability. In panel (b), in fact, the system undergoes orbital instability earlier than in panel (a), and the number of final planets is also smaller.

As for the first question posed in the beginning of this section, we find that in both simulations, one planet can acquire a thick H/He atmosphere from the disk, which may explain the origin of the observed low density super-Earths. Regarding the second question, we also observe that the acquisition of a massive atmosphere by one planet destabilizes closely spaced systems, leading to relatively well separated systems with fewer planets or even single-planet systems.

The properties (e.g., orbital separation and bulk density) of the final systems shown in Fig.~\ref{fig:t-a-env}(a) are reminiscent of some known planetary system. For example, two super-Earths were discovered in the Kepler-36 system, where the inner planet would be a rocky planet without a thick atmosphere, and the outer one would possess a thick atmosphere. However, the typical properties of the observed close-in super-Earths are unlikely to be reproduced. This is because the systems are tightly packed near the edge before the acquisition of atmospheres (see $t=10^5$ yr in Fig.~\ref{fig:t-a-env}), so that systems separated more than 2:1 resonances, which have been observed in super-Earth systems (see Fig.~\ref{fig:p-n}(a)), can hardly be reproduced by simulations for Model~1.

\citet{lee_etal14} point out that super-Earths with a mass of $10 M_\oplus$ tend to undergo runaway gas accretion, thus becoming gas giant planets. We do not observe this phenomenon in the results of Figs.~\ref{fig:t-a-env}(a) and (b), even though the core mass is high. This is because the massive planets move inside the disk inner edge, where there is no gas, and cease envelope accretion.

\section{Discussion and conclusions}
\label{sec:discussion}

We have re-examined in situ formation of close-in super-Earths with improved simulations, in which the effects of the disk of gas are considered. The simulations are started with small embryos and planetesimals. We find that the accretion of planets is extremely rapid owing to the large amount of solid material and short orbital periods. Thus, it is not correct to ignore the effects of the gas disk for the investigation of in situ formation of close-in super-Earths. 

We performed ten runs of simulation for our fiducial model and found the following. 1) The orbital architecture of resultant systems is very compact near the disk inner edge. 2) The eccentricity of super-Earths is small because planets can be stable after gas depletion. If they undergo orbital instability, the eccentricity is not highly excited. 3) The masses of planets monotonically decrease when increasing the semimajor axis. These characteristics are not consistent with observed close-in super-Earths. In fact, the cumulative observed distributions of period ratios of adjacent pairs and of eccentricities are statistically different from those we produce.

We have also investigated orbital evolution including the accretion process of primitive atmospheres onto the super-Earths. The results show that close-in super-Earths that formed in situ can acquire a thick H/He atmosphere in which the planets stop envelope accretion when they migrate inside the disk's inner edge.

Interestingly, if type I migration is neglected (but the eccentricity damping is included), the results match the observations much better. However, no mechanism capable of suppressing type I migration over the whole inner disk has ever been found. Recent studies have shown that MRI-driven disk winds, in which gas material is blown away from the surface of the disk, can alter the density profile of the gas disk, potentially slowing down or even reversing the migration of the protoplanets (e.g., \citealt{suzuki_inutsuka09}; \citealt{suzuki_etal10}; \citealt{ogihara_etal15}). This possibility, however, requires further investigation. Unless a mechanism for a global reduction of type I migration is demonstrated, our results imply that in situ accretion of close-in super Earth is unlikely.

\begin{acknowledgements}
We thank John Chambers for comments that helped us improve the manuscript. We also thank Yasunori Hori and Hiroki Harakawa for helpful comments. We thank the CRIMSON team, who manages the mesocentre of the OCA, on which most simulations were performed. Numerical computations were in part conducted on the general-purpose PC farm at CfCA of NAOJ. M.O. is supported by the JSPS Postdoctoral Fellowships for Research Abroad. A.M. and T.G. were supported by ANR, project number ANR-13--13-BS05-0003-01 projet MOJO(Modeling the Origin of JOvian planets).
\end{acknowledgements}

{}


\begin{thebibliography}{}
\bibitem[Adachi et al.(1976)]{adachi_etal76}
Adachi, I., Hayashi, C., \& Nakazawa, K. 1976,
Prog. Theor. Phys., 56, 1756
\bibitem[All\'egre et al.(2008)]{allegre_etal08}
All\'egre, C. J., Manh\'es, G. \& G\"opel, C. 2008,
Earth Planet. Sci. Lett., 267, 386
\bibitem[Bitsch \& Kley(2010)]{bitsch_kley10}
Bitsch, B., \& Kley, W. 2010,
\aap, 523, A30
\bibitem[Bitsch et al.(2014)]{bitsch_etal14}
Bitsch, B., Morbidelli, A., Lega, E., Kretke, K., \& Crida, A. 2014,
\aap, 570, A75
\bibitem[Chambers et al.(1996)]{chambers_etal96}
Chambers, J. E., Wetherill, Q. W., \& Boss, A. P. 1996,
\icarus, 119, 261
\bibitem[Chatterjee \& Tan(2014)]{chatterjee_tan14}
Chatterjee, S., \& Tan, J. 2014,
\apj, 780, 53
\bibitem[Chatterjee \& Tan(2015)]{chatterjee_tan15}
Chatterjee, S., \& Tan, J. 2015,
\apjl, 798, L32
\bibitem[Cossou et al.(2014)]{cossou_etal14}
Cossou, C., Raymond, S. N., Hersant, F., \& Pierens, A. 2014
\aap, 569, A56
\bibitem[Fabrycky et al.(2014)]{fabrycky_etal14}
Fabrycky, D. C., Lissauer, J. J., Ragozzine, D., et al. 2014,
\apj, 790, 146
\bibitem[Fendyke \& Nelson(2014)]{fendyke_nelson14}
Fendyke, S. M., \& Nelson, R. P. 2014,
\mnras, 437, 96
\bibitem[Hadden \& Lithwick(2014)]{hadden_lithwick14}
Hadden, A., \& Lithwick, Y. 2014,
\apj, 787, 80
\bibitem[Hansen \& Murray(2012)]{hansen_murray12}
Hansen, B. M., \& Murray, N. 2012,
\apj, 751, 158
\bibitem[Hansen \& Murray(2013)]{hansen_murray13}
Hansen, B. M., \& Murray, N. 2013,
\apj, 775, 53
\bibitem[Hayashi(1981)]{hayashi81}
Hayashi, C. 1981,
Prog. Theor. Phys. Suppl., 70, 35
\bibitem[Inamdar \& Schlichting(2015)]{inamdar_schlichting15}
Inamdar, N. K., \& Schlichteng, H. E. 2015,
\mnras, submitted
\bibitem[Ikoma \& Hori(2012)]{ikoma_hori12}
Ikoma, M., \& Hori, Y. 2012,
\apj, 753, 66
\bibitem[Ikoma et al.(2001)]{ikoma_etal01}
Ikoma, M., Emori, H., \& Nakazawa, K. 2001,
\apj, 553, 999
\bibitem[Johansen et al.(2014)]{johansen_etal14}
Johansen, A., Blum, J., Tanaka, H., Ormel, C., Bizzarro, M., \& Rickman, H. 2014,
in Beuther H., Klessen R., Dullemond C., Henning Th., eds, Protostars and Planets VI. Univ. Arizona Press, Tucson
\bibitem[Kokubo \& Ida(1998)]{kokubo_ida98}
Kokubo, E., \& Ida, S. 1998,
\icarus, 131, 171
\bibitem[Lee et al.(2014)]{lee_etal14}
Lee, E. J., Chiang, E., \& Ormel, C. W. 2014,
\apj, 797, 95
\bibitem[Lissauer et al.(2011a)]{lissauer_etal11a}
Lissauer, J. J. et al. 2011,
\nat, 470, 53
\bibitem[Lissauer et al.(2011b)]{lissauer_etal11b}
Lissauer, J. J. et al. 2011,
\apjs, 197, 8
\bibitem[Marcy et al.(2014)]{marcy_etal14}
Marcy, G. W., Weiss, L. M., Petigura, E. A., Isaacson, H., Howard, A. W., \& Buchhave, L. A. 2014,
PNAS, 111, 12655
\bibitem[Masset et al.(2006)]{masset_etal06}
Masset, F. S., D'Angelo, G., \& Kley, W. 2006,
\apj, 652, 730
\bibitem[Matsumoto et al.(2012)]{matsumoto_etal12}
Matsumoto, Y., Nagasawa, M., \& Ida, S. 2012,
\icarus, 221, 624
\bibitem[Mayor et al.(2009)]{mayor_etal09}
Mayor, M., et al. 2009, 
\aap, 493, 636
\bibitem[Morbidelli et al.(2008)]{morbidelli_etal08}
Morbidelli, A., Crida, A., Masset, F., \& Nelson, R. P. 2008,
\aap, 478, 929
\bibitem[Morbidelli et al.(2009)]{morbidelli_etal09}
Morbidelli, A., Bottke, W. F., Nesvorn\'y, D., \& Levison, H. F. 2009,
\icarus, 204, 558
\bibitem[Moorhead et al.(2011)]{moorhead_etal11}
Moorhead, A. V., Ford, E. B., Morehead, R. C., et al. 2011, 
\apjs, 197, 1
\bibitem[Mustill \& Wyatt(2011)]{mustill_wyatt11}
Mustill, A. J., \& Wyatt, M. C. 2011,
\mnras, 413, 554
\bibitem[O'Brien et al.(2006)]{obrien_etal06}
O'Brien, D. P., Morbidelli, A., \& Levison, H. F. 2006,
\icarus, 184, 39
\bibitem[Ogihara \& Ida(2009)]{ogihara_ida09}
Ogihara, M., \& Ida, S. 2009,
\apj, 699, 824
\bibitem[Ogihara \& Kobayashi(2013)]{ogihara_kobayashi13}
Ogihara, M., \& Kobayashi, H. 2013,
\apj, 775, L34
\bibitem[Ogihara et al.(2014)]{ogihara_etal14}
Ogihara, M., Kobayashi, H., \& Inutsuka, S. 2014,
\apj, 787, 172
\bibitem[Ogihara et al.(2015)]{ogihara_etal15}
Ogihara, M., Kobayashi, H., Inutsuka, S., \& Suzuki, T. K. 2015,
\aap, submitted
\bibitem[Paardekooper \& Mellema(2006)]{paardekooper_mellemal06}
Paardekooper, S. -J., \& Mellema, G. 2006,
\aap, 459, L17
\bibitem[Paardekooper et al.(2011)]{paardekooper_etal11}
Paardekooper, S. -J., Baruteau, C., \& Kley, W. 2011,
\mnras, 410, 293
\bibitem[Piso \& Youdin(2014)]{piso_youdin14}
Piso, A.-M., \& Youdin, A. N. 2014,
\apj, 786, 21
\bibitem[Pollack et al.(1996)]{pollack+1996}
Pollack, J. B., Hubickyj, O., Bodenheimer, P., Lissauer, J.J., Podolak, M., \& Greenzweig, Y. 1996,
\icarus, 124, 62
\bibitem[Schneider et al.(2011)]{schneider_etal11}
Schneider, J., Dedieu, C., Le Sidaner, P., Savalle, R., \& Zolotukhin, I. 2011,
\aap, 532, A79
\bibitem[Shen \& Turner(2008)]{shen_turner08}
Shen, Y., \& Turner, E. L. 2008,
\apj, 685, 553
\bibitem[Suzuki \& Inutsuka(2009)]{suzuki_inutsuka09}
Suzuki, T. K., \& Inutsuka, S. 2009,
\apj, 691, L49
\bibitem[Suzuki et al.(2010)]{suzuki_etal10}
Suzuki, T. K., Muto, T., \& Inutsuka, S. 2010,
\apj, 718, 1289
\bibitem[Tanaka et al.(2002)]{tanaka_etal02} 
Tanaka, H., Takeuchi, T., \& Ward, W. R. 2002,
\apj, 565, 1257
\bibitem[Tanaka \& Ward(2004)]{tanaka_ward04} 
Tanaka, H., \& Ward, W. R. 2004,
\apj, 602, 388
\bibitem[Wright et al.(2011)]{wright_etal11}
Wright, J. T., Fakhouri, O., Marcy, G. W., et al. 2011,
PASP, 123, 412
\bibitem[Tanigawa \& Ikoma(2007)]{tanigawa_ikoma07}
Tanigawa, T., \& Ikoma, M. 2007,
\apj, 667, 557
\bibitem[Ward(1986)]{ward86} 
Ward, W. R. 1986,
\icarus, 67, 164
\bibitem[Weidenschilling(1977)]{weidenschilling77}
Weidenschilling, S. J. 1977,
Ap\&SS, 51, 153
\bibitem[Wuchterl et al.(2000)]{wuchterl+2000}
Wuchterl, G., Guillot, T., \& Lissauer, J. J. 2000,
Protostars and Planets IV, 1081
\bibitem[Zakamska et al.(2011)]{zakamska_etal11}
Zakamska, N. L., Pan, M., \& Ford, E. B. 2011,
\mnras, 410, 1895
\end{thebibliography}

\end{document}